\documentclass[12pt]{iopart}
\pdfoutput=1
\usepackage{dcolumn}
\usepackage{bm}
\usepackage{graphicx}
\usepackage{color}
\usepackage{subfigure}
\usepackage{hyperref}
\usepackage{latexsym}
\usepackage{amsthm}
\usepackage{amssymb}
\usepackage{cite}
\DeclareGraphicsExtensions{.jpg,.pdf, .mps, .png, .eps, .ps, .EPS,.gif}

\begin{document}
\def\be{\begin{equation}}
\def\ee{\end{equation}}

\def\bc{\begin{center}}
\def\ec{\end{center}}
\def\bea{\begin{eqnarray}}
\def\eea{\end{eqnarray}}
\newcommand{\avg}[1]{\langle{#1}\rangle}
\newcommand{\Avg}[1]{\left\langle{#1}\right\rangle}

\def\ie{\textit{i.e.}}
\def\etal{\textit{et al.}}
\def\m{\vec{m}}
\def\G{\mathcal{G}}

\title[Quantum statistics in Network Geometry with Fractional Flavor]{Quantum statistics in Network Geometry with Fractional Flavor }

\author{Nicola Cinardi}
\address{Department of Physics and Astronomy {\it Ettore Majorana}, University of Catania,  via S. Sofia 64, Catania, Italy}

\author{Andrea Rapisarda}
\address{Department of Physics and Astronomy {\it Ettore Majorana}, University of Catania, and INFN, via S. Sofia 64, Catania, Italy\\
Complexity Science Hub Vienna, Austria}

\author{Ginestra Bianconi}

\address{
School of Mathematical Sciences, Queen Mary University of London, London, United Kingdom\\Alan Turing Institute, The British Library, London, United Kingdom\\}
\ead{g.bianconi@qmul.ac.uk}
\vspace{10pt}
\begin{indented}
\item[]
\end{indented}

\begin{abstract}
Growing network models have been shown to display emergent quantum statistics when nodes are associated to a fitness value describing the intrinsic ability of a node to acquire new links. Recently it has been shown that quantum statistics emerge also in  a growing simplicial complex model called Network Geometry with Flavor which allows for the description of many-body interactions between the nodes. This model depends on an external parameter called {\em flavor} that is responsible for the underlying topology of the simplicial complex. When the flavor takes the value $s=-1$ the $d$-dimensional simplicial complex is a manifold in which every $(d-1)$-dimensional face can only have an incidence number $n_{\alpha}\in\{0,1\}$. In this case the faces of the simplicial complex are naturally described by  the Bose-Einstein, Boltzmann and Fermi-Dirac distribution depending on their dimension. In this paper  we extend the study of Network Geometry with Flavor to fractional values of the flavor $s=-1/m$  in which every $(d-1)$-dimensional face can only have incidence number  $n_{\alpha}\in\{0,1,2,\dots, m\}$. We show that  in this case the statistical properties of the faces of the simplicial complex are described by the Bose-Einstein or the Fermi-Dirac distribution only. Finally we comment on the spectral properties of the networks constituting the underlying structure of the considered simplicial complexes. 
\end{abstract}
\section{Introduction}

Quantum statistics have been shown to  emerge spontaneously in the description of growing network models with fitness of the nodes \cite{Bose,BB,Fermi,Statistics,Rodgers,Chayes,Weight}.  In particular the Bianconi-Barab\'asi model \cite{Bose,BB} is a textbook \cite{NS} example of this phenomenon  in which the   non-equilibrium dynamics of a classical network is mathematically described by  quantum statistics. The implications of this mapping are profound. In particular the mapping of the Bianconi-Barab\'asi model with a Bose gas is able to predict a topological phase transition \cite{Bose,Godreche} in the network in which the dynamics of the networks is not stationary anymore but instead it is dominated by the sequence  of nodes with high fitness that arrive in the network and eventually become super-hubs. Interestingly this model has a large variety of applications ranging from the Internet \cite{Vespignani} and the WWW \cite{Adamic} and explains the basic mechanism beyond the winner-takes-all phenomena observed in such structures (like the emergence of super-hubs as Google and Facebook).
A symmetric generation of a  growing Cayley tree with fitness of nodes is instead described by the Fermi-Dirac distribution \cite{Fermi} and leads to the analytical description of Invasion Percolation on  these structures.

Recently these classical results of network theory have been related to the properties of growing simplicial complexes \cite{CQNM,Flavor,Hyperbolic,Weighted_SC,Polytopes}. A simplicial complex \cite{Perspective,Emergent,Equilibrium,Bassett,Lambiotte1} is a generalized network structure that allows the description of many-body interactions between a set of nodes. In particular simplicial complexes  are not only formed by nodes and links like networks but they are instead also formed by triangles, tetrahedra and so on.
Given that a simplicial complex is build by geometrical building blocks, simplicial complexes are natural structures to study network geometry. As such simplicial complexes have been widely used in quantum gravity to describe the discrete (or discretised) structure of space-time \cite{Loll,Oriti,Tensor,codello}.
In the last five years simplicial complexes are becoming increasingly popular to describe complex systems as well including collaboration networks, social networks, financial networks, nano-structures, and brain networks  \cite{Bassett,Lambiotte1,Vaccarino,Lambiotte2,Nanoparticles,Aste1,Petri,Latora,Ana,Ana2,Arenas,Ziff}. 
The Network Geometry with Flavor (NGF) \cite{Flavor,CQNM,Hyperbolic,Weighted_SC,Polytopes} is a non-equilibrium model of growing simplicial complexes with fitness that has been proposed to study emergent network geometry.
In fact the NGFs evolve thanks to purely combinatorial rules that make no use of any embedding space, but when the same length is attributed to each link of the simplicial complex they are able to generate structures with an emergent hyperbolic geometry \cite{Hyperbolic}.\\
The {\em flavor} $s$ of the NGF is a parameter that can change the topological nature of the simplicial complex and their evolution. For $s=-1$ the NGF is a manifold; for $s=0$ the network grows by uniform attachment of $d$-dimensional simplices on $(d-1)$-dimensional faces; finally for $s=1$ the network evolves according to a generalized preferential attachment rule  of of $d$-dimensional simplices on $(d-1)$-dimensional faces.

Interestingly NGFs have a stochastic topology that is described by quantum statistics \cite{Flavor,CQNM}.  In particular for $s=-1$ where we associate to the $(d-1)$-dimensional faces an incidence number $n_{\alpha}\in\{0,1\}$ we obtain that the $(d-1)$-dimensional faces are described by the Fermi-Dirac statistics. Moreover the lower dimensional faces are described by either the Boltzmann or the Bose-Einstein statistics. For instance in a NGF with flavor $s=-1$ and dimension  $d=3$ the statistical properties of the triangles, links and nodes of the simplicial complexes are described by the Fermi-Dirac, the Boltzmann and the Bose-Einstein statistics respectively.

In this paper we extend the study of this model to Network Geometry with Fractional Flavor $s=-1/m$ and $m>1$. In principle, since for these networks the incidence number of the $(d-1)$-faces is allowed to take only values $n_{\alpha}\in\{0,1,2\ldots, m\}$ we might expect to find that $(d-1)$-faces are described by  fractional statistics \cite{Gentile,Wilczek,Fractional}. Contrary to this naive expectation here we show that also in this case $(d-1)$-dimensional faces are described by the Fermi-Dirac statistics and that instead the main difference with the NGF with integer flavor $s=-1$ is that we do not find  any face described by the Boltzmann statistics.
This result sheds light on the effect that dimensionality and flavor have on the emergence of quantum statistics in NGFs. In particular while for integer flavor we must require $s=-1$ and $d\geq 3$ to observe the co-existence of the Fermi-Dirac and Bose-Einstein distribution describing the statistical properties of faces of different dimension $\delta$, for fractional flavor $s=-1/m$ (with $m>1$) we can have the coexistence of these two statistics already for dimension $d=2$.

\section{Simplicial complexes and their generalized degrees}

\subsection{Simplicial complexes}

\begin{figure}
\centering
  \includegraphics[width=0.9\columnwidth]{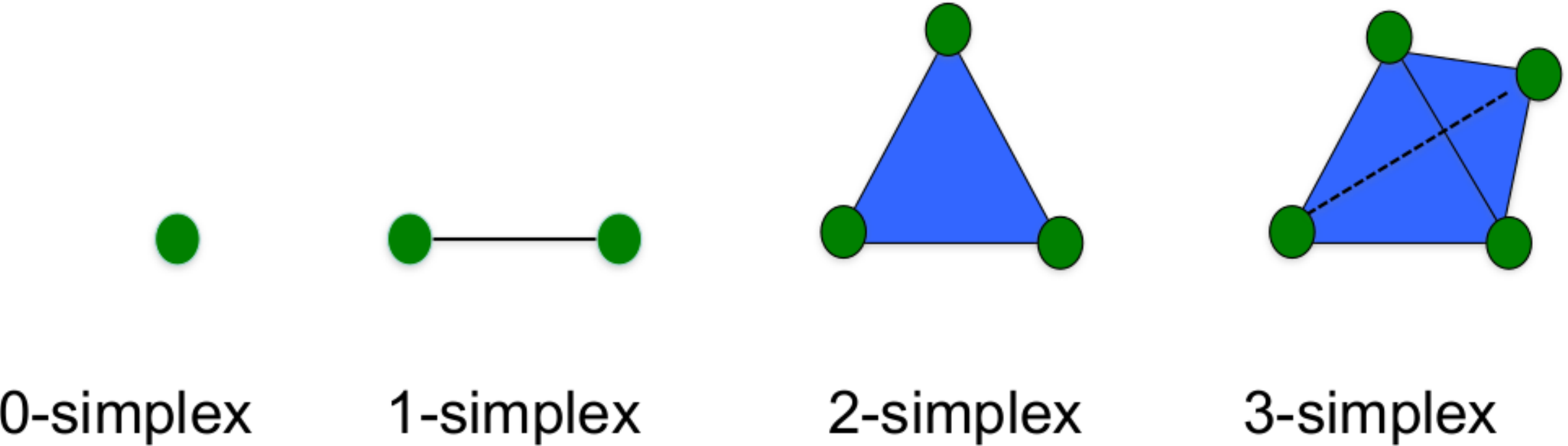}
\caption{Examples of $\delta$-dimensional simplices with $\delta=0,1,2,3$}
\label{fig:s1}      
\end{figure}

\begin{figure}
\centering
  \includegraphics[width=0.7\columnwidth]{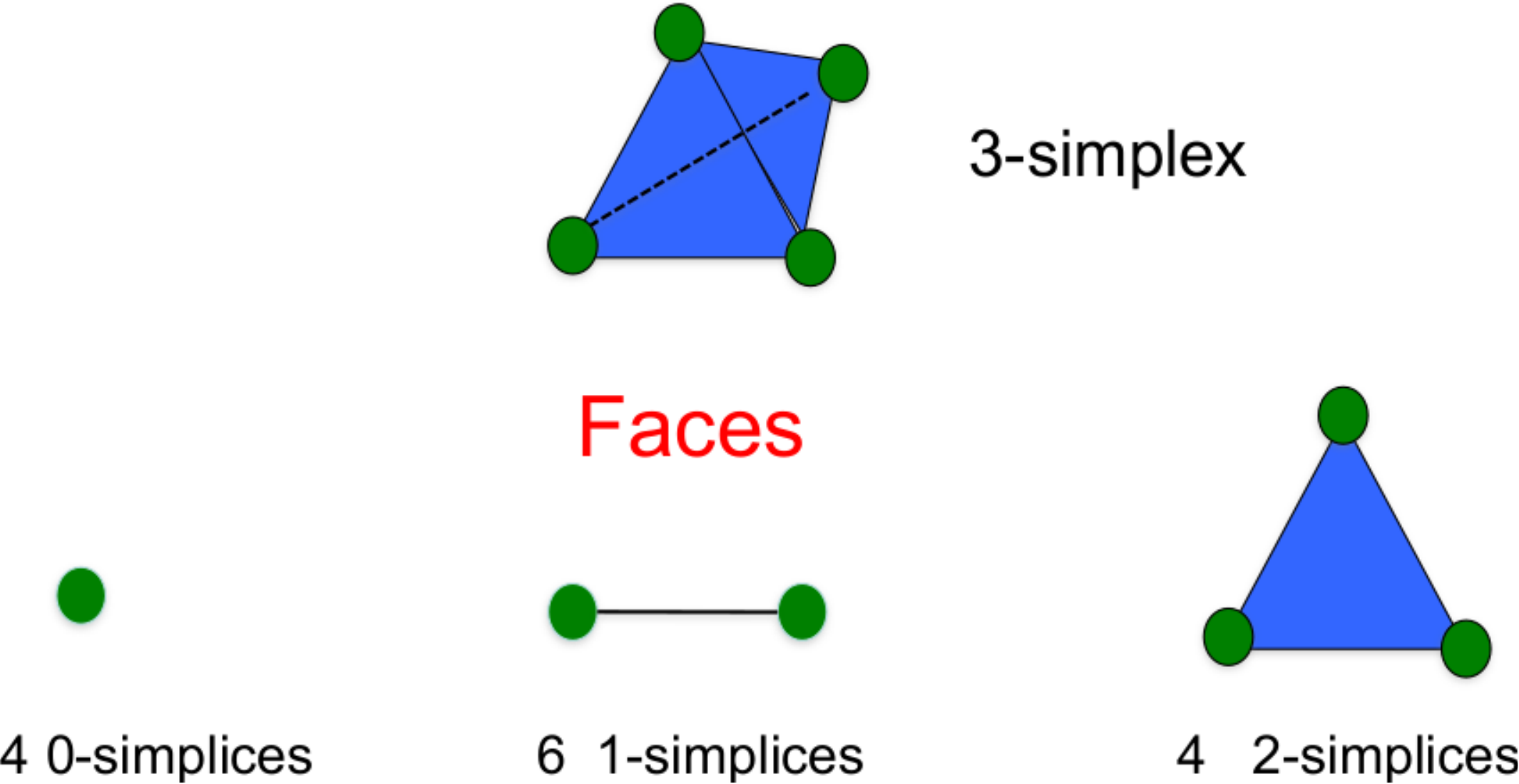}
\caption{The faces of a $\delta=3$ dimensional simplex (a tetrahedron) are shown and their relative number is indicated correspondingly.}
\label{fig:s2}      
\end{figure}

\begin{figure}
\centering
  \includegraphics[width=0.7\columnwidth]{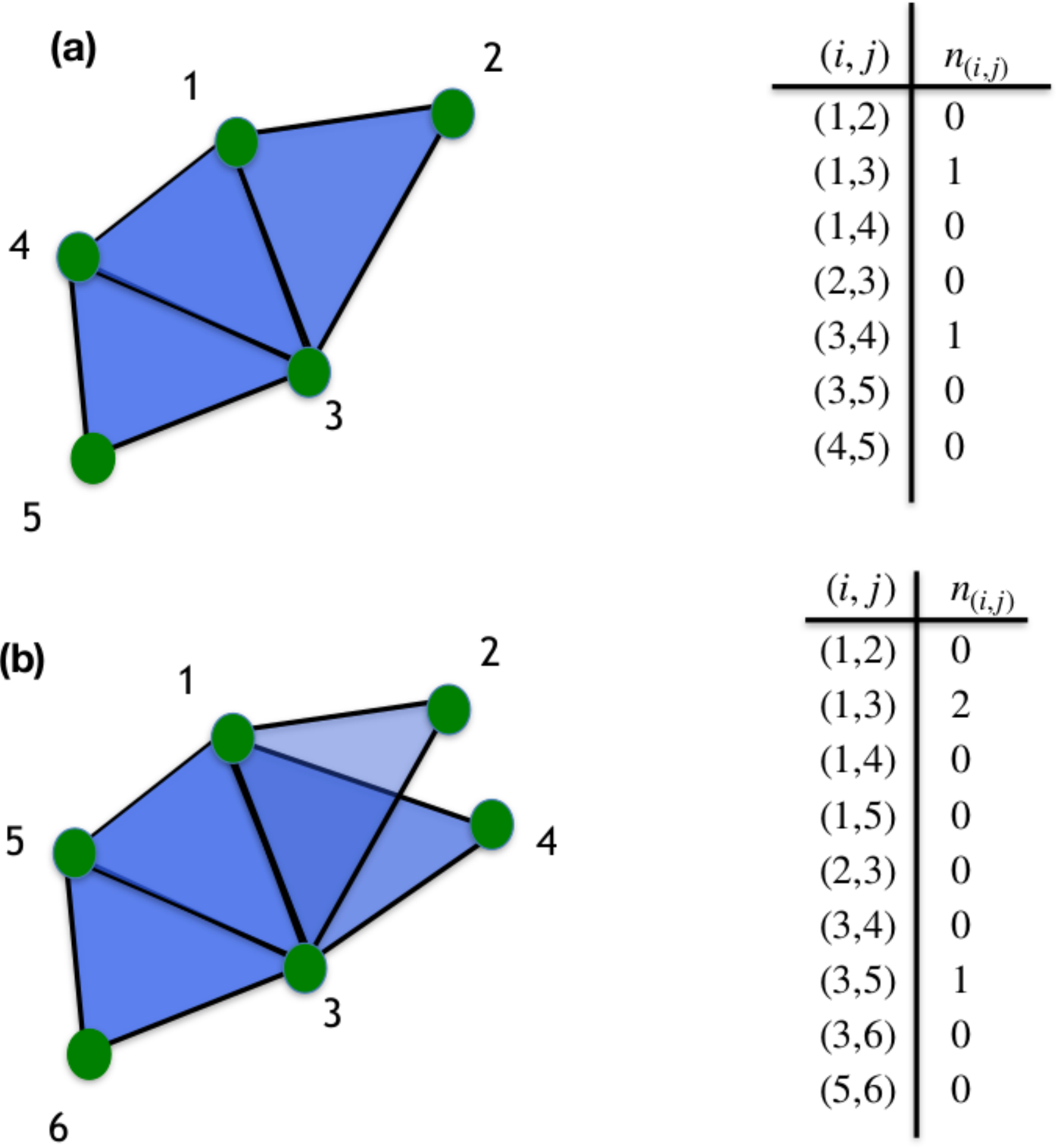}
\caption{Two examples of pure $2$-dimensional simplicial complexes are shown together with the list of the incidence number of their links. The simplicial complex in  panel (a) is a manifold as $n_{\alpha}\in \{0,1\}$ for all links $\alpha$ of the simplicial complex. The simplicial complex in panel (b) is not a manifold as $n_{(1,3)}=2$.}
\label{fig:s3}      
\end{figure}

A simplicial complex describes the many body interactions between a set of $N$ nodes. In particular a simplicial complex is formed by simplices glued along their faces.
A $\delta$-dimensional simplex is a set of $\delta+1$ nodes. Therefore a $0$-dimensional simplex is a node, a $1$-dimensional simplex is a link, a $2$-dimensional simplex is a triangle, and so on (see Figure \ref{fig:s1}).
A $\delta^{\prime}$-dimensional {\em face} $\alpha^{\prime}$ of a  $\delta$-dimensional simplex $\alpha$,  is a simplex formed by a subset of $\delta^{\prime}+1$ nodes of $\alpha$, i.e. $\alpha^{\prime}\subset \alpha$ (see Figure \ref{fig:s2}).
A $d$-dimensional {\em simplicial complex} ${\mathcal K}$ (see Figure $\ref{fig:s3}$ for examples) is formed by a set of simplices of dimensions $0\leq \delta\leq d$ (including at least a $d$-dimensional simplex) that obey the following two conditions:
\begin{itemize}
\item[(a)] if a simplex $\alpha$ belongs to the simplicial complex, i.e. $\alpha\in {\mathcal K}$ then also all its faces $\alpha^{\prime}\subset \alpha$ belong to the simplicial complex, i.e. $\alpha^{\prime}\in {\mathcal K}$;
\item[(b)] if two simplices $\alpha$ and $\alpha^{\prime}$ belong to the simplicial complex, i.e. $\alpha\in {\mathcal K}$ and $\alpha^{\prime}\in {\mathcal K}$, then either their intersection is the null set $\alpha\cap\alpha^{\prime}=\emptyset$
or their intersection belongs to the simplicial complex, i.e. $\alpha\cap\alpha^{\prime}\in{\mathcal K}$.
\end{itemize}

A $d$-dimensional simplicial complex is called {\em pure} if it is only formed by $d$-dimensional simplices and their faces (see Figure \ref{fig:s3} for examples).

From a simplicial complex it is always possible to extract a network called the {\em $1$-skeleton} by considering only the nodes and links of the simplicial complex.

In this paper we will focus on  pure $d$-dimensional simplicial complexes ${\mathcal K}$. In the following we will indicate with ${\cal Q}_{\delta}(N)$ the set of all possible $\delta$ simplices in a simplicial complex of $N$ nodes and with  ${\cal S}_{d, \delta}$  the set of all the $\delta$-dimensional faces of the pure $d$-dimensional simplicial complex ${\mathcal K}$.

Note that here we take  a purely topological \cite{top1} rather than a geometrical point of view. Therefore the nodes belonging to each simplex are not assigned a priori a position in an embedded space and the links of the simplices do not have an a priori defined length.  This is the ideal starting point for studying the problem of emergent hyperbolic network geometry \cite{Emergent} as one is interested on the minimal assumptions on the link lengths that allow the embedding on a space with a given sign of the  curvature.
\subsection{Generalized degrees }

 The topology of a pure $d$-dimensional simplicial complex ${\mathcal K}$  is fully specified by the adjacency tensor ${\bf a}$ of elements $a_{\alpha}$ with $\alpha\in{\cal Q}_{d}(N)$ given by 
 \bea
 a_{\alpha}=\left\{\begin{array}{cc}1&\mbox{if}\ \alpha\in{\mathcal K}, \\
 0&\mbox{otherwise}.\end{array}\right.
 \eea

The {\em generalized degree} $k_{d,\delta}(\alpha)$  \cite{Flavor,Equilibrium} of the $\delta$-face $\alpha$ is defined as the number of $d$-dimensional simplices  incident to it. 
Using the adjacency tensor we can evaluate  $k_{d,\delta}$ of a $\delta$-face $\alpha$ as
\bea
k_{d,\delta}(\alpha)=\sum_{\alpha' \in Q_{d}(N)| \alpha' \supset \alpha}a_{\alpha'}. 
\eea
Therefore,  in $d=2$, the generalized degree $k_{2,1}(\alpha)$ is the number of triangles incident to a link $\alpha$ while the generalized degree $k_{2,0}(\alpha)$ indicates the number of triangles incident to a node $\alpha$. 
Similarly in a  pure $d=3$ dimensional simplicial complex, the generalized degrees $k_{3,2}$, $k_{3,1}$ and $k_{3,0}$ indicate  the number of tetrahedra incident respectively to a triangular face, a link or a node. 
The generalized degrees of faces are not independent of the generalized degree of the simplices to which they belong \cite{Equilibrium}. In fact  the generalized degree of a $\delta-$face $\alpha$ is   related to the generalized degree of the $\delta'$-dimensional faces incident to it, with $\delta'>\delta$, by the simple combinatorial relation 
\bea
k_{d,\delta}(\alpha)=\frac{1}{\left(\begin{array}{c}d-\delta\\ \delta'-\delta\end{array}\right)}\sum_{\alpha'\in {\cal Q}_d (N)|\alpha'\supset \alpha}k_{d,\delta'}(\alpha').
\eea
Moreover, since every $d$-dimensional simplex belongs to $\left(\begin{array}{c}d+1\\ \delta+1\end{array}\right)$ $\delta$-dimensional faces, in a simplicial complex with $M$ $d$-dimensional simplices we have
\bea
\sum_{\alpha\in {\cal S}_{d,\delta}}k_{d,\delta}(\alpha)=\left(\begin{array}{c}d+1\\ \delta+1\end{array}\right)M.
\eea

\subsection{Incidence number}

The $(d-1)$-dimensional faces of a pure $d$-dimensional simplicial complex deserve some special attention. In particular to each $(d-1)$-dimensional face $\alpha$ we associate an {\em incidence number} $n_{\alpha}$ given by the number of incident $d$-dimensional simplices minus one, i.e.
\bea
n_{\alpha}=k_{d,d-1}(\alpha)-1.
\eea
Interestingly a simplicial complex can define a discrete $d$-dimensional manifold only if $n_{\alpha}\in \{0,1\}$, i.e.  a discrete $d$-dimensional manifold must have all its $(d-1)$-dimensional faces incident at most to two $d$-dimensional simplices. Therefore if $n_{\alpha}>1$ at least for one face $\alpha\in S_{d,d-1}$ then the simplicial complex is not a discrete manifold. In Figure $\ref{fig:s3}$ we show two examples of $2$-dimensional simplicial complexes (a manifold and a simplicial complex that is not a manifold) and the corresponding list of the incidence numbers of their links.

\section{Network Geometry with Flavor}

\subsection{Energy and fitness of the simplices}
In the Network Geometry with Flavor  each simplex $\alpha\in {\mathcal K}$ is associated to an energy $\epsilon_{\alpha}$ that does not change in time. The energy of a face describes its intrinsic and heterogeneous properties and has an important effect on the simplicial complex evolution.

The energy $\epsilon_i$ of node $i$ is   drawn randomly  from a given distribution $g(\epsilon)$.  
 To every  $\delta$-face  $\alpha\in {\cal S}_{d, \delta}$ with $0<\delta\leq d$ we associate an {\em energy} $\epsilon_{\alpha}$ given by the sum of the energy of the nodes that belong to the face  $\alpha$,
\bea
\epsilon_{\alpha}=\sum_{i \in \alpha}\epsilon_i.  
\label{ea}
\eea
Therefore, the energy of a  link is  given by the sum of energies of the two nodes that belong to it, the energy of a triangular face is given by  the sum of the energy of the three nodes belonging to it and so on. 
The energy $\epsilon_{(i,j)}$ of the generic   link $\alpha=(i,j)$ belonging to any given  triangle  of the NGF formed by the  nodes $i$, $j$ and $r$   satisfy the triangular inequality 
\bea
|\epsilon_{(i,r)}-\epsilon_{(j,r)}|\leq \epsilon_{(i,j)}\leq \epsilon_{(i,r)}+\epsilon_{(j,r)}.
\label{triangulard}
\eea 
This result remains valid for any permutation of the order of the nodes $i,j$ and $r$ belonging to the triangle. Therefore it is possible to consider  the energy of the link  as a possible candidate for the length of the link.

Finally to each simplex $\alpha\in \mathcal K$ we associate a {\em fitness } $\eta_{\alpha}$ given by 
\bea
\eta_{\alpha}=e^{-\beta\epsilon_\alpha},
\eea
where $\beta\leq 0$ is an external parameter of the model called {\em inverse temperature}.
If $\beta=0$ we have that $\eta_{\alpha}=1$ for every simplex $\alpha\in {\mathcal K}$, therefore every simplex has the same fitness independently of their differences in energy.
On the contrary when $\beta$ is large small differences in energy lead to large differences in the fitness of different simplices.

\subsection{Evolution of the Network Geometry with Flavor}
Network Geometry with Flavor (NGF) is a growing  model generating pure $d$-dimensional simplicial complexes. The stochastic evolution of NGF is determined by a parameter $s$ called the {\em flavor} and by the fitness of the simplices of the simplicial complex.
The evolution of the NGF obeys a simple iterative algorithm.\\
Initially at time $t = 1$ the simplicial complex is formed by a  single $d$-dimensional simplex. \\
At each time $t > 1$ we glue a $d$-dimensional simplex  to a $(d-1)$-face $\alpha$ chosen with probability 
\begin{equation}
\Pi_{d,d-1}(\alpha)=\frac{\eta_\alpha(1+sn_\alpha)}{Z^{[s]}},
\label{Pa}
\end{equation}
where $Z^{[s]}$ is called the {\em partition function} of the NGF and is given by 
\bea
Z^{[s]}(t)=\sum_{\alpha^{\prime}\in S_{d,d-1}} 
\eta_{\alpha^{\prime}}(1+sn_{\alpha^{\prime}}).
\eea
\subsection{Possible values of the flavor and their topological implications}

The Network Geometry with Flavor describes a growing simplicial complex that depends on the value of the flavor $s$.
Let us consider the attachment probability $\Pi_{d,d-1}(\alpha)$ for $\beta=0$ and the integer flavors $s\in \{-1,0,1\}$. In this case we have that the attachment probability satisfies 
\bea
\Pi_{d,d-1}(\alpha)\propto{(1+sn_\alpha)}=\left\{\begin{array}{lcc} 1-n_{\alpha} &\mbox{for}& s=-1,\\
1 &\mbox{for}& s=0,\\
k_{d,d-1}(\alpha)& \mbox{for} &s=1.
\end{array}\right.
\eea
Therefore the flavor $s=-1$ enforces the generation of a  manifold. In fact we have  $\Pi_{d,d-1}(\alpha)>0$ if $n_{\alpha}=0$ and $\Pi_{d,d-1}(\alpha)=0$ if $n_{\alpha}=1.$ Therefore  $n_{\alpha}\in\{0,1\}$ for every $(d-1)$-dimensional face $\alpha$ of the simplicial complex.
However in both cases $s=0$ and $s=1$  the incidence number can take any integer value $n_{\alpha}\geq 0$.
 The flavor $s=0$ corresponds to a uniform attachment of $d$-dimensional simplices of $(d-1)$-dimensional faces, while $s=1$ corresponds to a higher dimensional preferential attachment of $d$-dimensional simplices of $(d-1)$-dimensional faces.  

The NGF with integer flavor reduces to several known models for different values of the parameters $s,d$ and $\beta$.
For  $d=1,s=1,\beta=0$ the NGF reduces  to the Barab\'asi-Albert \cite{BA} model while for $d=1,s=1,\beta>0$ it reduces to the Bianconi-Barab\'asi model \cite{BB,Bose}.
For $d=2,s=0,\beta=0$ it reduces to the model proposed in Ref. \cite{Doro_link}
Finally for $d=3,s=-1,\beta=0$ it reduces to a random Apollonian network \cite{apollonian}.
 
 Values of the flavor $s$ different from the values $\{-1,0,1\}$ are also allowed as long as they lead to a suitable probability $\Pi_{d,d-1}(\alpha)\in [0,1]$ for every face $\alpha\in {\mathcal K}$.
Therefore positive values of the flavor $s\in \mathbb{R}^+$ are always allowed. In this case  via a rescaling of the attachment probability it is easy to show that NGFs  display a  stochastic topology with statistical properties equivalent to  NGF with flavor $s=1$.\\
For negative values of the flavor $s$ the requirement of observing a well defined attachment probability $\Pi_{d,d-1}(\alpha)\in [0,1]$ implies instead some restriction on the possible values of $s$. In particular if $s<0$, then $s$ should be of the form
\bea
s=-\frac{1}{m},
\eea
with $m\in \mathbb{N}$.
For such values of the flavor $s$  the incidence number of any $(d-1)$-dimensional face $\alpha$ can only take $m+1$ values, i.e.
\bea
n_{\alpha}\in \{0,1,2,\dots,m\}.
\eea
Therefore as long as $m>1$ NGF with $s=-1/m$ are not anymore manifolds, but they have a generalized degree of the $(d-1)$-dimensional faces bounded by $m+1$, i.e.
 \bea
k_{d,d-1}({\alpha})\in \{0,1,2,\dots,m+1\}.
\eea
This case, that we will call NGF with Fractional Flavor, is therefore expected to display statistical properties that are not equivalent to the ones observed for any of the integer flavors $s\in \{-1,0,1\}$. 
\begin{table}
\center
\caption{\label{table1} Distribution of generalized degrees of faces of dimension $\delta$ in a $d$-dimensional NGF of flavor $s$ at $\beta=0$.  For $d\geq 2\delta+2-s$ the power-law distributions are scale-free, i.e.  the second moment of the distribution diverges. }
\footnotesize
\begin{tabular}{@{}llll}
\hline
flavor &$s=-1$&$s=0$&$s=1$\\
\hline
$\delta=d-1$&Bimodal&Exponential&Power-law\\
\hline
$\delta=d-2$&Exponential&Power-law& Power-law\\
\hline
$\delta\leq d-3$&Power-law&Power-law& Power-law\\
\hline
\end{tabular}\\
\end{table}

\begin{table}
\center
\caption{\label{table2} The average  ${\Avg{k_{d, \delta}-1|\epsilon}}$  of the generalized degrees $k_{d,\delta}-1$ of $\delta$-faces with  energy $\epsilon$  in a $d$-dimensional NGF of flavor $s$ follows either the Fermi-Dirac, the Boltzmann or the Bose-Einstein statistics  depending on the values of the dimensions $d$ and $\delta$.  }
\footnotesize
\begin{tabular}{@{}llll}
\hline
flavor &$s=-1$&$s=0$&$s=1$\\
\hline
$\delta=d-1$&Fermi-Dirac &Boltzmann&Bose-Einstein\\
\hline
$\delta=d-2$&Boltzmann&Bose-Einstein& Bose-Einstein\\
\hline
$\delta\leq d-3$&Bose-Einstein&Bose-Einstein& Bose-Einstein\\
\hline
\end{tabular}
\end{table}
\section{Network Geometry with Integer Flavor}

The distribution $P_{d,\delta}(k)$ of generalized degrees $k_{d,\delta}=k$ of $\delta$-dimensional faces on the $d$-dimensional NGF has been  derived for integer flavors $s\in \{-1,0,1\}$ in Ref. \cite{Flavor}.
It has been found that for $\beta=0$  the generalized degree distribution $P_{d,\delta}(k)$  can follow  a bimodal, exponential or power-law distribution (see Table $\ref{table1}$) depending on the dimension $\delta$ and the flavor $s$ of the NGF.\\
For $\beta>0$ emergent  quantum statistics describe the statistical properties of NGFs  as long as the NGF has a stationary generalized degree distribution, i.e for sufficiently low value of $\beta\leq \beta_c$. Specifically it has been found in Ref. \cite{CQNM,Flavor}  that the average of the generalized degree minus one, $k_{d,\delta}-1$, over $\delta$-dimensional faces of energy $\epsilon$ can follow the Fermi-Dirac, the Boltzmann or the Bose-Einstein distribution depending on the dimension $\delta$ and the flavor $s$ of the NGF (see Table $\ref{table2}$). For instance in a NGF with $s=-1$ and $d=3$ the average of the generalized degree distribution minus one performed over faces of energy $\epsilon$ follows the Fermi-Dirac, the Boltzmann or the Bose-Einstein distribution for triangular faces, links and nodes respectively.

In the next section we will show how these statistical properties change for NGF with Fractional Flavor.

\begin{table}
\center
\caption{\label{table3} Distribution of generalized degrees of faces of dimension $\delta$ in a $d$-dimensional NGF of flavor $s$ at $\beta=0$.  Only for  $d-2\delta\geq 2+\frac{3}{m}$ the power-law distributions are scale-free, i.e.  the second moment of the distribution diverges. }
\footnotesize
\begin{tabular}{@{}ll}
\hline
flavor &$s=-1/m$\\
\hline
$\delta=d-1$&Bounded $k\leq m+1$\\
\hline
$\delta\leq d-2$&Power-law\\
\hline
\end{tabular}\\
\end{table}
\section{Network Geometry with Fractional Flavor and $\beta=0$}

\subsection{Main results}
In this section we will evaluate the generalized degree distribution  $P_{d,\delta}(k)$ of NGF with Fractional flavor $s=-1/m$ and $m>1$ for $\beta=0$. In particular we will show that differently from the cases $s=-1$ and $s=0$ the generalized degree distributions are never exponential.
In fact  for $s=-1/m$ and $m>1$ we obtain that the $(d-1)$-faces have a generalized degree distribution with bounded support with $k\leq m+1$ and the $\delta$-dimensional faces with $0\leq \delta<d-1$ have a generalized degree distribution which is power-law (see Table $\ref{table3}$).
In order to proof these results, in the following paragraph we first derive the generalized attachment probability. Subsequently   we derive the generalized degree distribution first using the mean-field approximation and  finally using the master equation approach providing exact asymptotic results.

\subsection{Attachment probability}
For fractional flavor $s=-\frac{1}{m}$ the attachment probability for $\beta=0$, given by Eq. (\ref{Pa}) can  also be expressed as  
\bea
\Pi_{d,d-1}({\alpha})=\frac{m-n_\alpha}{\tilde{Z}}=\frac{m+1-k_{d,d-1}(\alpha)}{\tilde{Z}},
\eea
where $\tilde{Z}$ is given by 
\bea
\tilde{Z}&=&\sum_{\alpha\in {\mathcal S}_{d,d-1}}(m-n_{\alpha})=\sum_{\alpha\in {\mathcal S}_{d,d-1}}(m+1-k_{d,d-1}({\alpha})).\nonumber \eea
Therefore the normalization constant $\tilde{Z}$ counts each $(d-1)$-dimensional face $\alpha$ with a degeneracy $m-n_{\alpha}\in\{0,1,2,\ldots, m\}$.\\
Since at time $t=1$ we have $d+1$ $(d-1)$-dimensional faces  with degeneracy $m$ we have that $\tilde{Z}(t=1)=m(d+1)$. Moreover at each time we add $d$ new  $(d-1)$-dimensional faces with degeneracy $m$ and we reduce the degeneracy of the $(d-1)$-faces $\alpha$ to which we add the new $d$-dimensional simplex by $1$. Therefore at each time $\tilde{Z}$ increases by $md-1$. It follows that the normalization constant $\tilde{Z}$ is given by 
\bea
\tilde{Z}&=&=(md-1)t+1+m\simeq (md-1)t,
\eea
where the last expression is valid for $t\gg1$.
The probability ${\Pi}_{d,\delta}(\alpha)$ that a new $d$-dimensional simplex is attached to a   $\delta\leq d-2$ dimensional face $\alpha$ is given by  
\bea
 {\Pi}_{d,\delta}(\alpha)= \sum_{\alpha'\in S_{d,d-1}|\alpha'\supset \alpha}\frac{m-n_{\alpha^{\prime}}}{\tilde{Z}}.
 \label{Pad}
\eea 
In order to calculate the numerator of this expression we make the following considerations.
If we assume that the face $\alpha$ has incidence number $n_{\alpha}=0$, then the numerator of Eq. (\ref{Pad}) is given by $d-\delta$.
In fact  every $\delta$-dimensional face $\alpha$ with $\delta<d-1$ and generalized degree $k_{d,\delta}(\alpha)=1$ is incident to 
\bea
{{d-\delta}\choose{d-\delta-1}}=d-\delta
\eea
$(d-1)$-dimensional faces with degeneracy $m$. This follows from the fact that its incident  $(d-1)$-dimensional faces must contain  $d-\delta-1$ nodes that do not belong to the face $\alpha$. These nodes should  chosen among the  $(d-\delta)$ nodes of the single $d$-dimensional simplex that contains the face $\alpha$ and are external to face $\alpha$.
Following a similar argument it is easy to check that at  each time we add a $d$-dimensional simplex to the $\delta$-dimensional face the number of  $(d-1)$-dimensional  faces with degeneracy $m$  increases by   
\bea
{{d-\delta-1}\choose{d-\delta-2}}=d-\delta-1
\eea
and additionally we reduce the degeneracy of the $(d-1)$-dimensional face to which we attach the new $d$-dimensional simplex by one.
Therefore for $t\gg1 $ we have
\bea
{\Pi}_{d,\delta}(\alpha)\simeq \left\{\begin{array}{lcc}\frac{m+1-k_{d,\delta}(\alpha)}{(md-1)t} & \mbox{for}& \delta = d-1,\\ &&\\
 \frac{[m(d-\delta-1)-1]k_{d,\delta}(\alpha)+m+1}{(md-1)t} & \mbox{for}& \delta\leq d-2. \end{array}\right.
\eea
Finally given the above expression we can express the probability $ \tilde{\Pi}_{d,\delta}(k)$ that a new $d$-dimensional simplex is attached to a $\delta$-dimensional face $\alpha$ with generalized degree $k_{d,\delta}(\alpha)=k$ as
\bea
 \tilde{\Pi}_{d,\delta}(k)\simeq \left\{\begin{array}{lcc}\frac{m+1-k}{(md-1)t} & \mbox{for}& \delta = d-1,\\ &&\\
 \frac{[m(d-\delta-1)-1]k+m+1}{(md-1)t} & \mbox{for}& \delta\leq d-2. \end{array}\right.
\eea
\subsection{Mean-field approach}
In order to find an approximated generalized degree distribution we can consider the popular mean field approach \cite{Doro_book}. In this case we assume that the generalized degrees can be approximated by their average over different network realizations and that they evolve by  a deterministic equation
\begin{equation}
\frac{dk_{d,\delta}(\alpha)}{dt}={\Pi}_{d,\delta}(\alpha).
\label{mf}
\end{equation}
Let us consider separately the case in which $\delta=d-1$ and the case $\delta\leq d-2$.
The mean-field equation for the generalized degree of  $\delta=d-1$  dimensional faces  is given by  
 \begin{equation}
\frac{dk_{d,d-1}}{dt}=\frac{m+1-k_{d,d-1}}{(md-1)t},
\end{equation}
 with initial condition $k_{d,d-1}(t_{\alpha})=1$. It follows that in the mean-field approximation the generalized degree of a $(d-1)$-face added at time $t_{\alpha}$ evolves in time as \bea
k_{d,d-1}(t)=m+1-m\left(\frac{t_\alpha}{t}\right)^{1/(md-1)}.
\eea
If follows that, as expected,  the generalized degrees of the $(d-1)$-dimensional faces are bounded and  asymptotically in time saturate to the the value $k_{d,d-1}=m+1$.
In order to derive the generalized degree distribution in the mean-field approximation we calculate the probability ${\mathcal P}(k_{d,\delta}>k)$. This is given by 
\bea
{\mathcal P}(k_{d,d-1}>k)&=&{\mathcal P}\left(t_\alpha<\left(\frac{m+1-k}{m}\right)^{md-1}t\right)\nonumber \\
&&=\left(\frac{m+1-k}{m}\right)^{md-1}.
\eea
Therefore in the mean-field approximation the generalized degree distribution is given by 
\bea
\tilde{P}_{d,d-1}(k)=-\frac{{\mathcal P}(k_{d,d-1}>k)}{dk}=\frac{md-1}{m}\left(\frac{m+1-k}{m}\right)^{md-2},
\eea
 valid for $1\leq k\leq m.$
Let us now consider the mean-field equation for  the $\delta \leq d-2$ dimensional faces. This equation reads
\begin{equation}
\frac{dk_{d,d-1}}{dt}=\frac{[m(d-\delta-1)-1]k_{d,\delta}+m+1}{(md-1)t}
\end{equation}
 with initial condition $k_{d,d-1}(t_{\alpha})=1$. Therefore in the mean field approximation the generalized degree of a $\delta$-face added at time $t_{\alpha}$ grows in time as
\bea
k_{d,d-1}(t)&=&\left(1+1\right)\left(\frac{t}{t_\alpha}\right)^{\frac{m(d-\delta-1)-1}{md-1}}-a
\eea 
where for convenience we have  defined the constant $a$ as
\bea
a=\frac{m+1}{m(d-\delta-1)-1}.
\label{a}
\eea
Given this definition we can express the generalized degree distribution of $\delta\leq d-2$ faces in the mean field approximation. In fact we have that the cumulative generalized degree distribution is given by 
\bea
{\mathcal P}(k_{d,\delta}>k)&=&{\mathcal P}\left(t_\alpha<\left(\frac{1+a}{k+a}\right)^{\frac{md-1}{m(d-\delta-1)-1}}t\right)\nonumber \\
&&=\left(\frac{1+a}{k+a}\right)^{\frac{md-1}{m(d-\delta-1)-1}}.
\eea
By differentiating this expression we find the generalized degree distribution of $\delta\leq d-2$ faces in the mean field approximation is given by 
\bea
\tilde{P}_{d,\delta}(k)&=&-\frac{{\mathcal P}(k_{d,\delta}>k)}{dk}\nonumber \\&=&\frac{md-1}{m(d-\delta)}\left(\frac{1+a}{k+a}\right)^{\frac{md-1}{m(d-\delta-1)-1}+1}.
\eea
Therefore we find in this approximation that generalized degrees have a bounded distribution for 
$\delta\leq d-1$ faces and a power-law distribution for $\delta\leq d-2$ faces.

\begin{figure}
  \includegraphics[width=0.9\columnwidth]{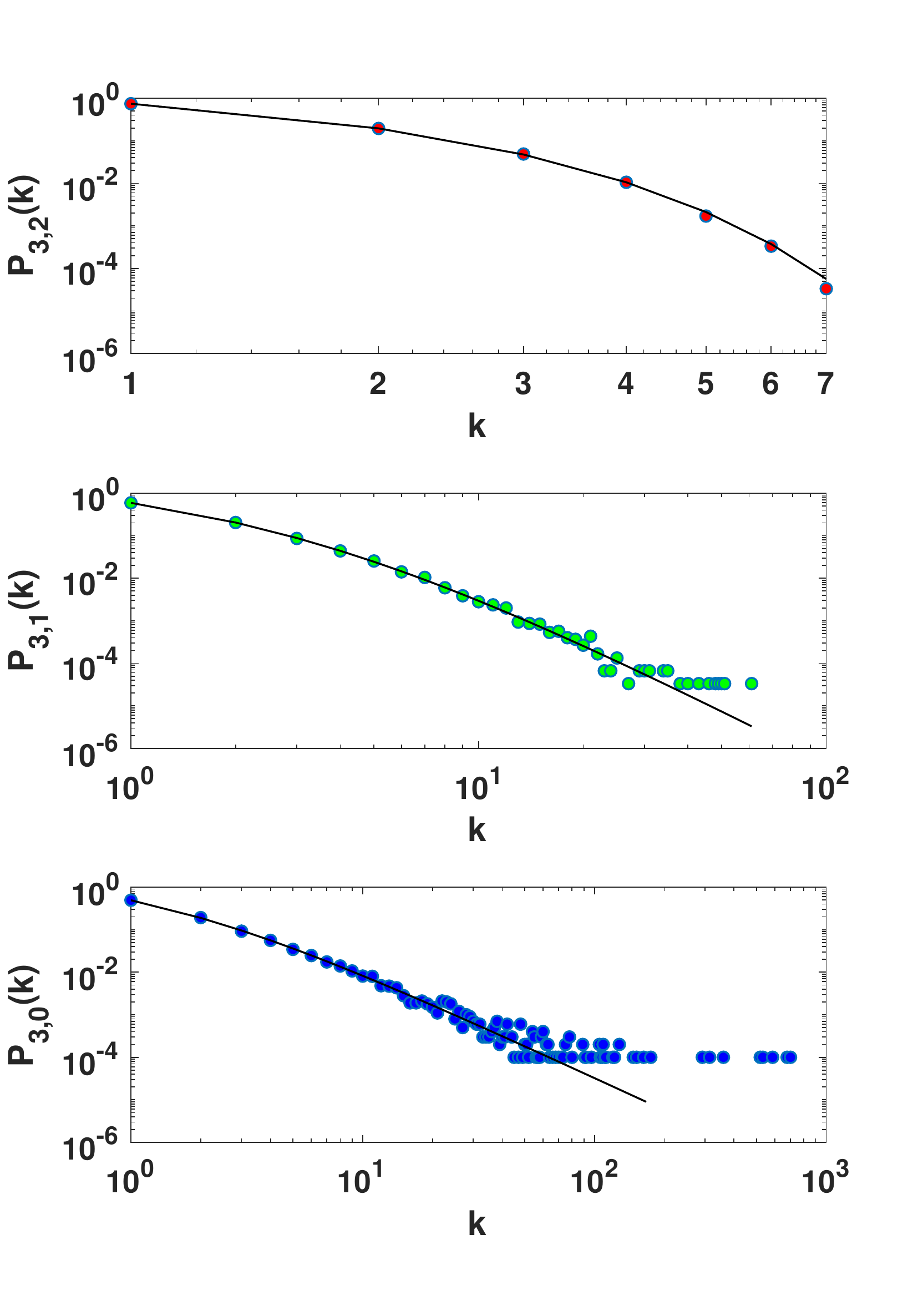}
\caption{Generalized degree distribution $P_{d,\delta}(k)$ of nodes ($\delta=0$), links ($\delta=1$) and triangles ($\delta=2$) of  a NGF with $N=5000$ nodes, flavor $s=-1/6$, dimension $d=3$ and inverse temperature $\beta=0$. The symbols indicate the results of simulations, the solid lines indicate the theoretical predictions obtained using the master equation approach.}
\label{fig:1}      
\end{figure}

\subsection{Master equation approach}

The mean-field approach gives only approximate results for the generalized degree distribution. In order to get the exact asymptotic results we need to consider the master equation approach \cite{Doro_book}.
The master equation describes the evolution of the average number $N_{d,\delta}^{t}(k)$ of $\delta$-dimensional faces that at time $t$ have generalized degree $k$ in a $d$-dimensional NGF with flavor $s=-\frac{1}{m}$.
We notice that at each time we add  $$m_{d,\delta}= {{d}\choose{\delta}}$$ number of $\delta$-faces of generalized degree $k_{d,\delta} = 1$ and the average number of $\delta$-faces of generalized degree $k_{d,\delta} =k$ increases by $${\Pi}_{d,\delta}(k-1)N_{d,\delta}^{t}(k-1)$$ if $k>1$ and decreases by $${\Pi}_{d,\delta}(k)N_{d,\delta}^{t}(k).$$ Therefore the master equation reads
\bea
N_{d,\delta}^{t+1}(k)&=&N_{d,\delta}^{t}(k) + {\Pi}_{d,\delta}(k-1)N_{d,\delta}^{t}(k-1)(1-\delta_{k,1})\nonumber \\ &&-{\Pi}_{d,\delta}(k)N_{d,\delta}^{t}(k)+m_{d,\delta}\delta_{k,1}.
\label{ME}
\eea
 For large network sizes when  $t\gg1$ the average number of $\delta$-dimensional faces is given by  \bea
 N_{d,\delta}^{t}(k)\simeq m_{d,\delta}tP_{d,\delta}(k)
 \label{Nasym}
 \eea
 where $P_{d,\delta}(k)$ is the generalized degree distribution of the $\delta$-dimensional faces.
Inserting this asymptotic expression in the Eq. $(\ref{ME})$ we can derive the generalized degree distribution as explained in the following by distinguishing between the case in which $\delta=d-1$ and the case in which $\delta\leq d-2$.

For the $\delta=d-1$ dimensional faces, by using the expression 
 \bea
 \Pi_{d,d-1}(k)=\frac{m+1-k}{(md-1)t},
 \eea 
 and the asymptotic scaling of $N_{d,\delta}^{t}(k)$ given by Eq. $(\ref{Nasym})$  the master equation can be re-written in terms of the generalized degree distribution obtaining
 \bea
P_{d,d-1}(k)&=& \frac{m+2-k}{(md-1)}P_{d,d-1}(k-1)(1-\delta_{k,1})+\nonumber\\ &&-\frac{m+1-k}{(md-1)}P_{d,d-1}(k)+\delta_{k,1}.
 \eea
 obtaining 
\begin{equation}
P_{d,d-1}(k)=
\frac{md-1}{m(d+1)-1}\frac{\Gamma(m+1)}{\Gamma(md+m-1)}\frac{\Gamma(md+m-k)}{\Gamma(m-k+2)},\nonumber
\end{equation}
valid for $1\leq k\leq m.$

For the $\delta \leq d-2$ dimensional faces by using the expression 
 \bea
 \Pi_{d,\delta}(k)=\frac{[m(d-\delta-1)-1]k+m+1}{(md-1)t},
 \eea 
 and the asymptotic scaling of $ N_{d,\delta}^{t}(k)$ given by Eq. $(\ref{Nasym})$  the master equation can be re-written in terms of the generalized degree distribution obtaining
 \bea
\hspace*{-5mm}P_{d,\delta}(k)&=&\frac{[m(d-\delta-1)-1]k+m+1}{md-1}P_{d,\delta}(k-1)(1-\delta_{k,1})\nonumber \\ \hspace*{-5mm}&&-\frac{[m(d-\delta-1)-1]k+m+1}{md-1}P_{d,\delta}(k)+\delta_{k,1}.\nonumber
 \eea
 This latter recursive equation has explicit solution
  \bea
P_{d,\delta}(k)&=&
\frac{md-1}{m(2d-\delta)-1}\frac{\Gamma\left(2+\frac{m(d+1)}{m(d-\delta-1)-1}\right)}{\Gamma\left(1+\frac{m+1}{m(d-\delta-1)-1}\right)}\nonumber \\
&&\times\frac{\Gamma\left(k+\frac{m+1}{m(d-\delta-1)-1}\right)}{\Gamma\left(k+1+\frac{m(d+1)}{m(d-\delta-1)-1}\right)}. 
\eea
This distribution for large $k\gg1 $ decays as a power-law 
\bea
P_{d,\delta}(k)&\simeq& k^{-\gamma_{d,\delta}}
\eea
with 
\bea
\gamma_{d.\delta}=1+\frac{md-1}{m(d-\delta-1)-1}.
\eea
These distributions are therefore scale-free, i.e. $\gamma\leq 3$ if and only if 
\bea
d-2\delta\geq 2+\frac{3}{m}. 
\eea
These theoretical predictions show that the generalized degree distribution is indeed bounded for $\delta=d-1$ dimensional faces and power-law for $\delta\leq d-2$ dimensions. 
 As expected these results perfectly match the simulation results providing exact asymptotic expression for the generalized degree distribution $P_{d,\delta}(k)$ for NGF with fractional flavor $s=-\frac{1}{m}$ (see Figure $\ref{fig:1}$).

\section{Network Geometry with Fractional Flavor and $\beta>0$}

\subsection{Main results}

Quantum statistics have been shown to characterize the statistical properties of NGF with integer flavor $s\in \{-1,0,1\}$. In particular $d$-dimensional NGFs with flavor $s=-1$ have an average degree of $\delta$ faces with energy $\epsilon$ described by the Fermi-Dirac (for $\delta=d-1$), the Boltzmann (for $\delta=d-2$) and the Bose-Einstein distribution (for $\delta\leq d-3$). On the contrary on NGF with flavor $s=0$ the average degree of $\delta$ faces with energy $\epsilon$ can be only described by the Boltzmann and the Bose-Einstein  distribution. Finally in NGF with flavor $s=1$ all the faces, independently of their dimension $\delta$, have an average degree  described only by the Bose-Einstein distribution.
Interestingly if we consider integer flavors $s\in\{-1,0,1\}$ the Fermi-Dirac distribution emerges as the natural distribution characterizing the statistical properties of $\delta=d-1$ faces only if the flavor is given by $s=-1$, which corresponds to the case in which the incidence number $n_{\alpha}$ fo the $\delta=d-1$ faces can only take the values $n_{\alpha}\in\{0,1\}$.
This suggests a relation between the emergence of the Fermi-Dirac statistics and the constraint imposed by the flavor $s=-1$ on the possible values of the incidence number.
It is therefore interesting to investigate the properties of NGF with factional flavor $s=-\frac{1}{m}$ in which the  only allowed values  of the incidence number are $n_{\alpha}\in\{0,1,\ldots, m\}$.
In principle one could expect that in this case the statistical properties of the generalized degree of the NGF would be   described by generalized quantum statistics such as the Gentile statistics \cite{Gentile} or the anyons statistics \cite{Wilczek,Fractional}.
To our surprise instead the result of our calculations has revealed that in NGF with fractional flavor $s=-\frac{1}{m}$ and $m>1$ does not display any fractional statistics (see Table \ref{table4}).
If we compare the results obtained for to the NGFs with   fractional flavor $s=-\frac{1}{m}$ and $m>1$ to the results obtained for NGF with $s=-1$  we observe that
\begin{itemize} 
\item  the average degree of $(d-1)$-dimensional faces with energy $\epsilon$ still remain described by the Fermi-Dirac statistics also if the  incidence number of $\delta=d-1$ faces can take values $n_{\alpha}\in\{0,1,\ldots, m\}$ with $m>1$;
\item
the average degree of $\delta=(d-2)$ dimensional faces with energy $\epsilon$ is already characterized by the Bose-Einstein statistics and not by the Boltzmann statistics;
\item
 the average degree of $\delta< d-2$-dimensional faces with energy $\epsilon$ is characterized by the Bose-Einstein statistics like in the case $s=-1$.
\end{itemize}

In particular these results imply that when we consider the fractional flavor  $s=-\frac{1}{m}$ and $m>1$ we have that already for $d=2$ dimensional NGF we can observe the coexistence of faces with statistical properties described respectively by the Fermi-Dirac and Bose-Einstein distribution while for observing the co-existence of these two statistics in NGF with flavor $s=-1$ we should have dimension $d\geq 3$.

\begin{table}
\center
\caption{\label{table4}The average  ${\Avg{k_{d, \delta}-1|\epsilon}}$  of the generalized degrees $k_{d,\delta}$ of $\delta$-faces with  energy $\epsilon$ minus one in a $d$-dimensional NGF of flavor $s=-\frac{1}{m}$ follows either the Fermi-Dirac or the Bose-Einstein statistics  depending on the values of the dimensions $d$ and $\delta$.  }
\footnotesize
\begin{tabular}{@{}llll}
\hline
&s=-1/m\\
\hline
$\delta=d-1$&Fermi-Dirac \\
\hline
$\delta\leq d-2$&Bose-Einstein\\
\hline
\end{tabular}
\end{table}
\subsection{Attachment probability and chemical potentials}

When $\beta>0$ the generalized degree distribution of the NGF can be solved by extending the self-consistent approach proposed for solving the Bianconi-Barab\'asi model \cite{BB,Bose} which constitutes the NGF model for $s=1$ and $d=1$.
In this approach it is assumed that the statistical properties of the NGF reach a steady state and that it is possible to define  suitable parameters $\mu_{d,\delta}$ called {\em chemical potentials}.
In particular the chemical potential $\mu_{d,d-1}$ of the $\delta=d-1$ faces is defined as 
\bea
e^{\beta \mu_{d,d-1}}=\lim_{t\to \infty}\frac{t}{mZ^{[s]}},
\eea
while the chemical potential  $\mu_{d,\delta}$ of the $\delta<d-1$ faces  is defined as 
\bea
e^{\beta \mu_{d,\delta}}=\lim_{t\to \infty}\Avg{\frac{\sum_{\alpha^{\prime}\in S_{d,d-1}|\alpha\subset \alpha^{\prime}}e^{-\beta (\epsilon_{\alpha'}-\epsilon_{\alpha})}(1+sn_{\alpha^{\prime}})t}{Z^{[s]}(a+k_{d,\delta}(\alpha))}},\nonumber
\eea
where here the average is done over $\delta$-dimensional faces ${\alpha\in S_{d,\delta}}$.
In both cases it is assumed that if the network evolution reaches a stationary state, then the chemical potential is self-averaging, i.e. it does not depend on the specific network realization of the NGF over which the limit $t\to \infty$ is performed.
As long as the chemical potentials are well determined and self-averaging quantities the attachment probabilities can be expressed in terms of the chemical potentials ,and  it can be easily shown that the probability $ {\Pi}_{d,\delta}(\alpha)$ that a new $d$-dimensional simplex is attached to a new $\delta$-dimensional face $\alpha$ is given by 
\bea
 {\Pi}_{d,\delta}(\alpha)\simeq \left\{\begin{array}{lcc}e^{-\beta(\epsilon_{\alpha}-\mu_{d,d-1})}\frac{m+1-k_{d,\delta}(\alpha)}{t} & \mbox{for}& \delta = d-1,\\\
e^{-\beta(\epsilon_{\alpha}-\mu_{d,d-1})} \frac{k_{d,\delta}(\alpha)+a}{t} & \mbox{for}& \delta\leq d-2, \end{array}\right.
\eea
where  $a$ is given by Eq. ($\ref{a}$). From this it follows that the probability $\tilde{\Pi}_{d,\delta}(k)$ that a new $d$-dimensional simplex is attached to a $\delta$-face with generalized degree $k_{d,\delta}(\alpha)=k$ is given by 
\bea
 \tilde{\Pi}_{d,\delta}(k)\simeq \left\{\begin{array}{lcc}e^{-\beta(\epsilon_{\alpha}-\mu_{d,d-1})}\frac{m+1-k}{t} & \mbox{for}& \delta\leq d-1,\\ &&\\
e^{-\beta(\epsilon_{\alpha}-\mu_{d,d-1})} \frac{k+a}{t} & \mbox{for}& \delta\leq d-2. \end{array}\right.
\label{pkb}
\eea
\begin{figure*}
  \includegraphics[width=0.9\columnwidth]{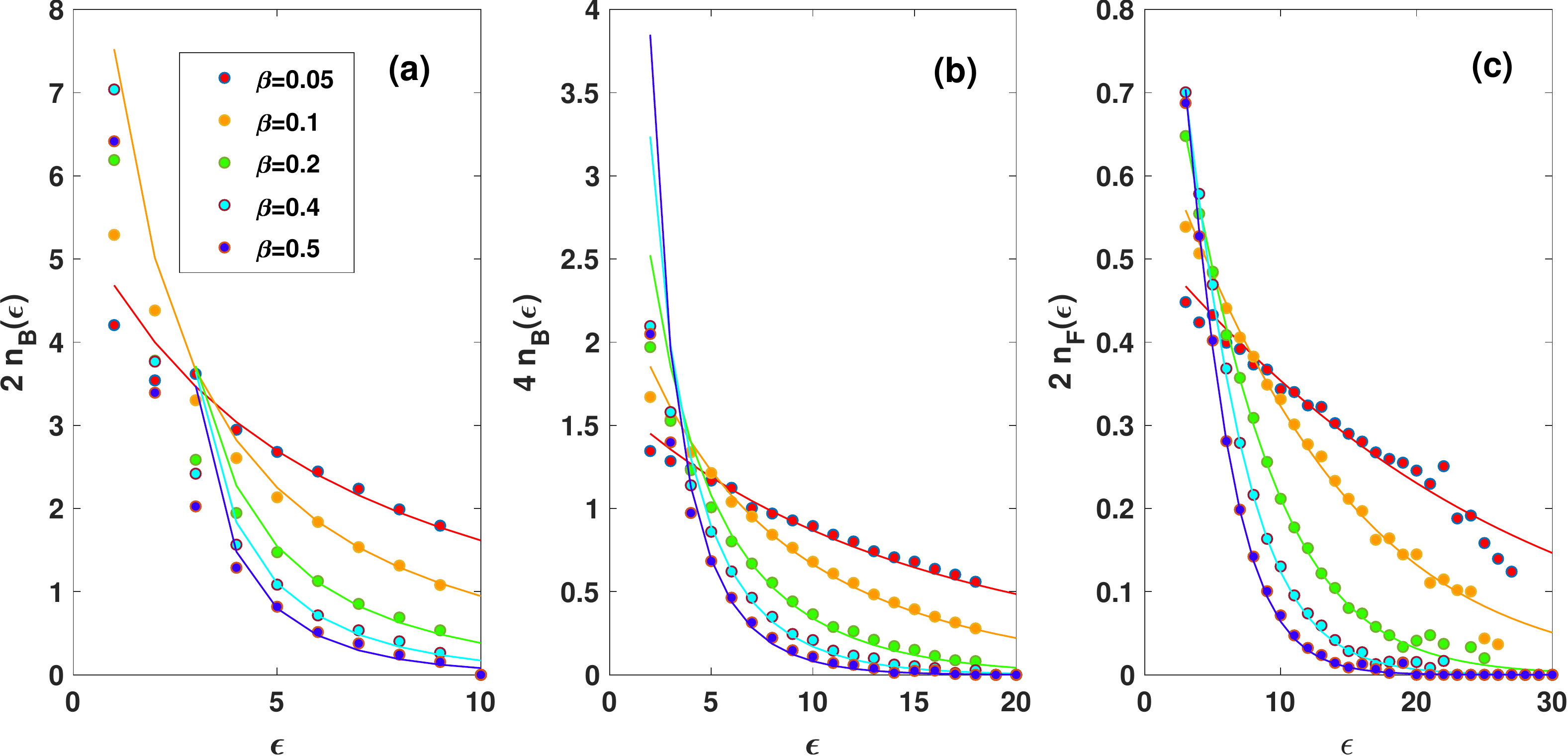}
\caption{Average generalized degree minus one, over faces of energy $\epsilon$ and dimension $\delta=0$ (panel a) $\delta=1$ (panel b) or $\delta=2$ (panel c) for $d=3$ dimensional NGF with fractional flavor $s=-1/2$ are plotted for different values of $\beta$ (symbols) and compared to the  theoretical expectations (Fermi-Dirac and Bose-Einstein statistics). The simulations are performed for  NGF with $N=3000$ nodes. The data are averaged over $50$ NGF realizations.}
\label{fig:2}      
\end{figure*}
\subsection{Mean-field approach}
Let us consider first the results that can be obtained within the mean field approximation.
As mentioned before for the case $\beta=0$ in the mean-field approximation we neglect the fluctuations and we consider a deterministic evolution of the generalized degree that is assumed to be equal to the average generalized degree over different NGF realizations. 
Let us consider separately the case in which the cases $\delta=d-1$ and  $\delta\leq d-2$ .
By assuming that the chemical potential $\mu_{d,d-1}$   is well defined, and using Eq. (\ref{pkb}) for the attachment probability{\tiny }, the mean-field equation (Eq. (\ref{mf})) for generalized degree $k_{d,d-1}(\alpha)=k_{d,d-1}$ of the generic $(d-1)$-face $\alpha$ with energy $\epsilon_{\alpha}=\epsilon$ can be written explicitly as 
 \begin{equation}
\frac{dk_{d,d-1}}{dt}=\frac{e^{-\beta(\epsilon-\mu_{d,d-1})}(m+1-k_{d,d-1})}{t},
\end{equation}
with initial condition $k_{d,d-1}(t_{\alpha})=1$.
This equation has solution
\bea
k_{d,d-1}(t|\epsilon)=m+1-m\left(\frac{t_\alpha}{t}\right)^{e^{-\beta(\epsilon-\mu_{d,d-1})}},
\eea
which  like in the case $\beta=0$ clearly implies that the generalized degree of the $(d-1)$-dimensional faces is bounded. Interestingly in the mean-field approximation we can evaluate the average of the generalized degrees  minus one over faces with energy $\epsilon$ getting
\begin{equation}
\Avg{k_{d,d-1}-1|\epsilon}=\frac{m}{e^{\beta(\epsilon_{\alpha}-\mu_{d,d-1})}+1}=mn_F(\epsilon).
\end{equation}
Therefore this quantity is proportional to the Fermi-Dirac distribution $n_F(\epsilon)$ with chemical potential $\mu_{d,d-1}$. Interestingly, as we will show in the next paragraph this result is exact, in fact it is a result that concerns the average of the generalized degrees and therefore is not affected by the mean-field approximation.
However the generalized degree distribution of $d-1$ faces that can be derived from the mean-field approach is  instead an approximation. By proceeding similarly to the case $\beta=0$ we obtain that in the mean field approximation the probability $\tilde{P}_{d,d-1}(k|\epsilon)$ that a $(d-1)$-dimensional face with energy $\epsilon$ has generalized degree $k_{d,d-1}(\alpha)=k$ is given  by    
\bea
\hspace*{-3mm}\tilde{P}_{d,d-1}(k|\epsilon)=\frac{1}{m}e^{\beta(\epsilon-\mu_{d,\delta})}\left(\frac{m+1-k}{m}\right)^{e^{\beta(\epsilon-\mu_{d,\delta})}-1}.
\eea
We can proceed similarly  for the $\delta \leq d-2$ dimensional faces. In particular in this case  the mean-field equations read
\begin{equation}
\frac{dk_{d,\delta}}{dt}=\frac{e^{-\beta(\epsilon-\mu_{d,\delta})}(k_{d,\delta}+a)}{t},
\end{equation}
with initial condition $k_{d,\delta}(t_{\alpha})=1$.
Here we have assumed that the chemical potential $\mu_{d,\delta}$ is well defined and we have used Eq. (\ref{pkb}) for the attachment probability ${\Pi}_{d,\delta}$. The above mean-field equations have the solution 
\bea
k_{d,\delta}(t|\epsilon)&=&\left(1+a\right)\left(\frac{t}{t_\alpha}\right)^{e^{-\beta(\epsilon_{\alpha}-\mu_{d,\delta})}}-a.
\eea 
By using this expression it is possible to calculate the average of the generalized degree minus one over faces of energy $\epsilon$ finding
\bea
\Avg{k_{d,\delta}-1|\epsilon}= A_{\delta}\frac{1}{e^{\beta(\epsilon_{\alpha}-\mu_{d,\delta})}-1}=A_{\delta}n_B(\epsilon),
\eea
where 
\bea
A_{\delta}=1+a=\frac{m(d-\delta)}{m(d-\delta-1)-1}.
\eea
Therefore we find that $\delta\leq d-2$ dimensional faces of energy $\epsilon$ have statistical properties described by the Bose-Einstein distribution $n_B(\epsilon)$ with chemical potential $\mu_{d,\delta}$.
However in the mean-field approximation the derived generalized degree distribution is not exact but approximated. Proceeding as in the previous case we find  that in the mean field approximation the 
 probability $\tilde{P}_{d,d-1}(k|\epsilon)$ that a $(d-1)$-dimensional face with energy $\epsilon$ has generalized degree $k_{d,d-1}(\alpha)=k$ is given by   
\bea
\tilde{P}_{d,\delta}(k|\epsilon)=e^{\beta(\epsilon-\mu_{d,\delta})}(1+a)\left(\frac{1+a}{k+a}\right)^{e^{\beta(\epsilon-\mu_{d,\delta})}+1}
\eea

\subsection{Master equation approach}

For $\beta>0$ we can find the exact asymptotic result for the generalized degree distribution of faces of given energy $\epsilon$.
The master equation from which we start is written for the number $N_{d,\delta}^t(k|\epsilon)$ of $\delta$-dimensional faces with energy $\epsilon$ and reads
\bea
N_{d,\delta}^{t+1}(k|\epsilon)&=&N_{d,\delta}^{t}(k|\epsilon) + {\Pi}_{d,\delta}(k-1)N_{d,\delta}^{t}(k-1|\epsilon)(1-\delta_{k,1})\nonumber \\ &&-{\Pi}_{d,\delta}(k)N_{d,\delta}^{t}(k|\epsilon)+\rho_{d,\delta}(\epsilon)\delta_{k,1}.\nonumber
\label{MEB}
\eea
where $ {\Pi}_{d,\delta}(k-1)$ is given by Eq. (\ref{pkb}) and where $\rho_{d,\delta}(\epsilon)$ indicates the density of new faces with energy $\epsilon$ that we add at time $t$.
 For large network sizes when  $t\gg1$ the average number of $\delta$-dimensional faces with energy $\epsilon$ is given by  \bea
 N_{d,\delta}^{t}(k|\epsilon)\simeq \rho_{d,\delta}(\epsilon)tP_{d,\delta}(k|\epsilon).
 \label{Nasymb}
 \eea
 Inserting this asymptotic expression we get the exact asymptotic result for the generalized degree distribution $P_{d,\delta}(k|\epsilon)$ of $\delta$-dimensional faces with energy $\epsilon$.
 Specifically in the case $\delta=d-1$ we obtain the bounded distribution
\bea
P_{d,d-1}(k|\epsilon)&=&
\frac{e^{\beta(\epsilon-\mu_{d,d-1})}}{(e^{\beta(\epsilon-\mu_{d,d-1})}+m)}\frac{\Gamma\left(m+1\right)}{\Gamma\left(m+e^{\beta(\epsilon-\mu_{d,d-1})}\right)}  \nonumber \\
&&\times \frac{\Gamma\left(m-k+1+e^{\beta(\epsilon-\mu_{d,d-1})}\right)}{\Gamma\left(m-k+2\right)},
\eea
for $1\leq k\leq m+1$.
For $\delta\leq d-2$ we obtain instead the power-law distribution
\bea
P_{d,\delta}(k|\epsilon)&=&\frac{e^{\beta(\epsilon-\mu_{d,\delta})}[m(d-\delta-1)-1]}
{e^{\beta(\epsilon-\mu_{d,\delta})}[m(d-\delta-1)-1]+m(d-\delta)} \nonumber \\
&&\times \frac{\Gamma\left(2+{a}+e^{\beta(\epsilon-\mu_{d,\delta})}\right)}
{\Gamma\left(1+{a}\right)} \nonumber \\
&&\times\frac{\Gamma\left(k+{a}\right)}{\Gamma\left(k+1+{a}+e^{\beta(\epsilon-\mu_{d,\delta})}\right)}.
\eea
Therefore for $k\gg1$ the generalized degree distribution of $\delta$-dimensional  faces with energy $\epsilon$ decays as a power-law with an energy dependent power-law exponent $\gamma(\epsilon)$, i.e.
\bea
P_{d,\delta}(k|\epsilon)\simeq k^{-\gamma(\epsilon)}
\eea
with  
\bea
\gamma(\epsilon)=1+e^{\beta(\epsilon-\mu_{d,\delta})}.
\eea
Having the exact asymptotic results of the generalised degree distribution $P_{d,\delta}(k|\epsilon)$ valid as long as the chemical potentials $\mu_{d,\delta}$ are well defined, we can perform the average over all $\delta$-faces with energy $\epsilon$, i.e. 
\bea
\Avg{k_{d,\delta}-1|\epsilon}=\sum_{k}(k-1)P_{k,\delta}(k|\epsilon).
\eea
In this result, we obtain total agreement with the mean-field results, i.e.
\bea
\Avg{k_{d,d-1}-1|\epsilon } =   m n_F(\epsilon)    && \mbox{for} \quad \delta=d-1, \\
\Avg{k_{d,\delta}-1|\epsilon } =  A_{\delta}n_B(\epsilon) && \mbox{for} \quad \delta\leq d-2.
\eea
Therefore the  generalized degree minus one averaged over faces of energy $\epsilon$  is proportional to the Fermi-Dirac distribution for $\delta=d-1$ while it is proportional to the Bose-Einstein distribution for faces of dimension $\delta\leq d-2$.  
In Figure $\ref{fig:2}$ we compare the simulation results with the theoretical predictions showing very good agreement as long as the inverse temperature $\beta$ is sufficiently low.
For higher values of $\beta$ the system does not reach a stationary state and the description of this phase transition is beyond the scope of this work.
\begin{figure}[htb!]
\begin{center}
\includegraphics[width=0.95\columnwidth]{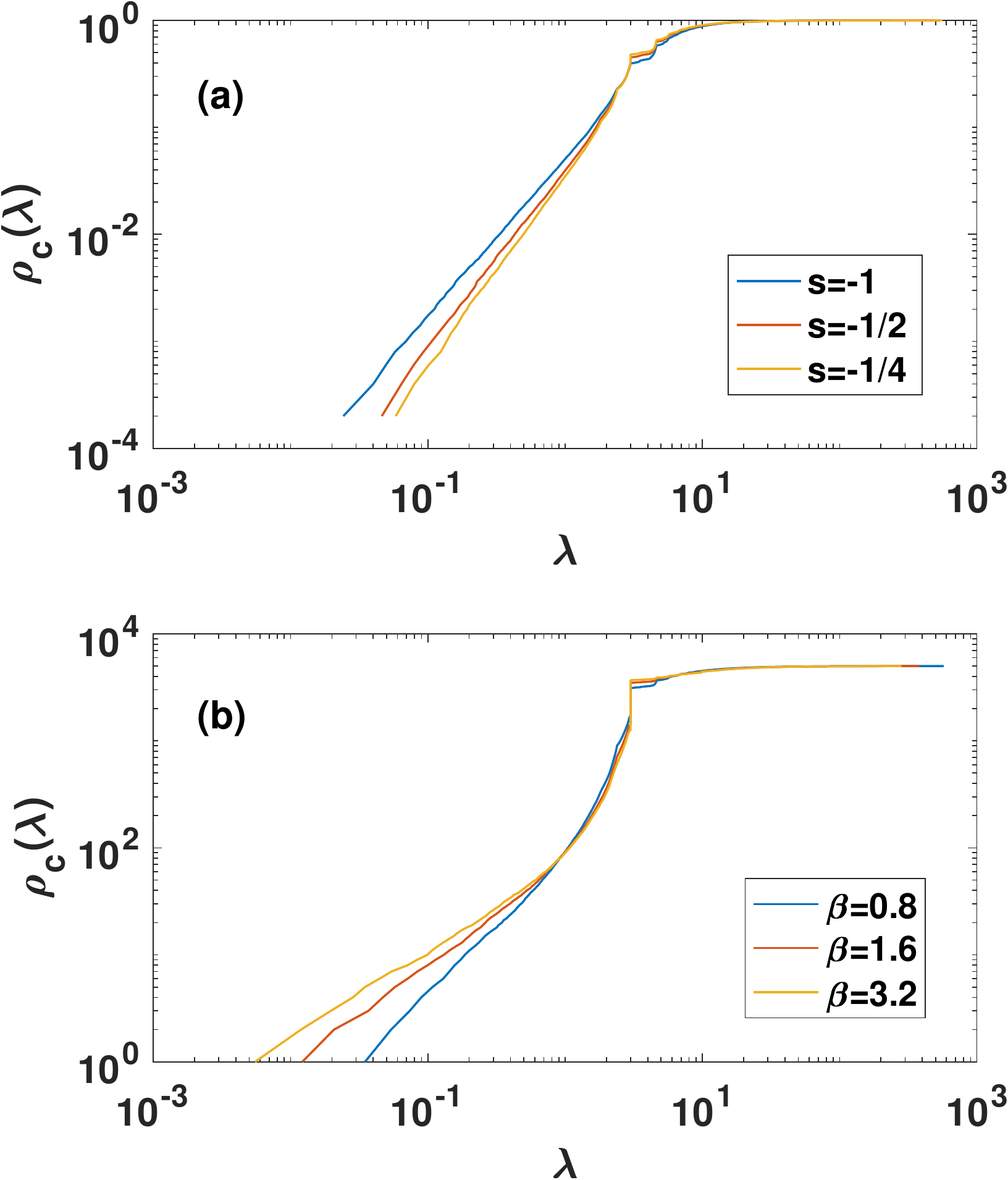} 
\caption{Cumulative distribution $\rho_c(\lambda)$ of the eigenvalues of the Laplacian for NGFs of dimension $d=3$ formed by   $N=5000$ nodes. In panel (a) we show the cumulative distribution $\rho_c(\lambda)$ as a function of $s$ for $\beta=0$. In panel (b) we show the cumulative distribution $\rho_c(\lambda)$ as a function of $\beta$ for $s=-1/2$. Every spectrum is averaged over   10 realizations of the NGFs.}
\label{fig:spectral_beta}
\end{center}
\end{figure}
\section{Spectral properties of the NGF with Fractional Flavor}
The spectral dimension \cite{Toulouse} determines the properties of a diffusion process defined on the  1-skeleton of the NGF, i.e. the network constructed by starting from the simplicial complex by considering exclusively its nodes and links.
Given the Laplacian matrix ${\bf L}$ with elements defined as 
\bea
L_{ij}=K_i\delta_{ij}-a_{ij}
\eea
where ${\bf a}$ is the adjacency matrix of the network,  $K_i$ indicates the degree of the generic node $i$, and the density of eigenvalues $\rho(\lambda)$ for $\lambda\ll1 $  obeys the power-law scaling 
\bea
\rho(\lambda)\simeq \lambda^{d_S/2-1}
\eea
we say that the network has {\em spectral dimension} $d_S$.
Note that in this case the cumulative density of eigenvalues $\rho_c(\lambda)$ obeys the scaling 
\bea
\rho_c(\lambda)\simeq \lambda^{d_S/2}
\eea
for $\lambda\ll1$.
The NGFs with integer flavor $s\in\{-1,0,1\}$ have been shown to display a finite spectral dimension \cite{Ana,Ana2,Polytopes}. Therefore it is interesting to investigate here how the spectral dimension changes for NGF with fractional flavor. 
By calculating numerically the spectrum of large NGF we  found that for  $\beta=0$ the spectral dimension $d_S$ of the NGF with flavor $s=-\frac{1}{m}$ is an increasing  function of $m$. Therefore it  achieves its smallest value for $s=-1$ and increases as $m$ increases  (see Figure $\ref{fig:spectral_beta}$a).
Moreover the spectral properties of the NGF changes also with $\beta$. In particular numerical results indicate that  the spectral dimension $d_S$ decreases as the inverse temperature $\beta$
 increases (see Figure $\ref{fig:spectral_beta}$b).

\section{Conclusions}

In conclusion here we have extended the model Network Geometry with Flavor to fractional negative values of the flavor $s=-\frac{1}{m}$. This choice of parameters enforces the condition that each $(d-1)$-dimensional face of the pure $d$-dimensional simplicial complexes generated by the model is incident at most to $m+1$ $d$-dimensional simplices.
For the limiting case $m=1$ this model generates discrete manifolds where $(d-1)$-dimensional faces have incidence numbers $n_{\alpha}\in \{0,1\}$.  For $m>1$ instead the simplicial complexes generated by the model are not anymore manifolds   and have incidence number $n_{\alpha}\in \{0,1,2,\ldots, m\}$.
In previous studies it has been shown that NGF displays emergent quantum statistics. In particular  the generalized degrees of $\delta$ faces for energy $\epsilon$ in  the NGFs with $s=-1$ can be simply related to the Fermi-Dirac, Boltzmann and Bose-Einstein statistics depending on the dimensionality $\delta$. 
This result implies that for a NGF in $d=3$ with $s=-1$ the triangles, the links, and the nodes of given energy $\epsilon$ have an average generalized degree  minus one given by the Fermi-Dirac, the Boltzmann and the Bose-Einstein statistics respectively.
Here we show that when we consider NGF with flavor $s=-\frac{1}{m}$ we still observe different statistics as a function of the dimensionality of the faces but the only two types of statistics emerging are the Fermi-Dirac and Bose-Einstein statistics as long as the NGF evolution reaches a steady state.
This implies that for $d=2$ we observe links and nodes  of energy $\epsilon$ whose average generalized degree follows the Fermi-Dirac and Bose-Einstein distribution respectively.
Therefore already in $d=2$ we observe the co-existence of the two quantum statistics determining properties of faces of different dimension.
Finally in this paper we have also numerically analysed the spectral properties of NGF with fractional flavor showing how the spectral dimension of the 1-skeleton of NGF changes as a function of $m$ and $\beta$.

The proposed NGF model with fractional flavor can be used to model real simplicial complexes in which nodes have some intrinsic features that can be associated with their fitness. As generalized network structures with metadata are increasingly available we believe that this is a very promising possible application of our modeling framework.
Moreover the NGF with fractional flavor can be used as  well controlled artificial models in which to test dynamical processes defined on simplicial complexes, such as topological percolation \cite{Ziff}, synchronization \cite{Ana,Ana2,Arenas} or social contagion \cite{Latora}.

\section*{Acknowledgements}
G. B. acknowledges interesting discussions with  L. Smolin and  R. Sorkin.
G.B. was partially supported by the  Perimeter Institute for Theoretical Physics (PI). The PI is supported by the Government of Canada through Industry Canada and by the Province of Ontario through the Ministry of Research and Innovation. A.R. acknowledges financial support by  the project {\it Linea di intervento 2} of the  Department of Physics and Astronomy {\it Ettore Majorana} of the University of Catania.

\end{document}